\documentstyle[12pt]{article}

\begin{document}
\rightline{CAT/96-02}
\begin{center}
{\Large{\bf Non-Abelian Antisymmetric-Vector Coupling\\ 
from Self-Interaction}}\\ [7mm]
{\bf Adel Khoudeir}\footnote { e-mail: adel@ciens.ula.ve}\\ 
{\it Centro de Astrof\'{\i}sica Te\'orica, Departamento de F\'{\i}sica,
Facultad de Ciencias, Universidad de los Andes, M\'erida, 5101,
Venezuela.}\\[4mm] 

{\bf Abstract}
\end{center}

A non-abelian coupling between antisymmetric fields and Yang-Mills
fields 
proposed by Freedman and Townsend several years ago is derived using the 
self-interaction mechanism.

\section{INTRODUCTION}
Abelian second-rank antisymmetric fields \cite{og} play an essential 
role in strings and supergravity theories and have been extensively 
studied in the last decades \cite{kr} \cite{cs} \cite{ft} \cite{others}. 
In free theories they describe massless and spinless particles and
appear 
in many 
contexts, for instance, arising as mediators of the interaction between
open 
strings with charged particles \cite{kr} and in ten dimensions, coupling 
with the Chern-Simons 3-form  to achieve an elegant unification of 
Yang-Mills and supergravity \cite{vcm}. In particular the Cremmer-Sherk 
theory \cite{cs} has received considerable attention \cite{all}
\cite{dem} 
due to the fact that 
the coupling between the abelian antisymmetric field and a Maxwellian 
field through a topological $BF$ term leads to massive propagations
which 
are compatible with gauge invariances. Moreover, Allen, et. al.
\cite{all} 
have shown unitarity and renormalizability of the Cremmer-Sherk theory. 
This fact motivates the non-abelian generalization of the model 
and several attempts 
have been proposed \cite{lah}. Simultaneously, other alternatives for 
non-abelian massive 
vector bosons without the presence of Higgs field have been proposed in 
the last year \cite{vani}. 

The non-abelian extension of antisymmetric theories was achieved by 
Freedman and Townsend \cite{ft} starting from a first-order formulation 
where the antisymmetric field $B_{mn}^a$ and an auxiliary vector
potential 
are independent variables. It is worth recalling that the non-abelian 
generalization of the abelian S-duality theory \cite{witten} is a
Freedman-
Townsend theory \cite{yloz}. In their work, Freedman and Townsend
proposed 
the non-abelian generalization of the Cremmer-Sherk theory.
In this letter, starting from an appropiate first-order formulation for
the 
Cremmer-Sherk theory, we will derive the non-abelian generalization 
using the self-interaction mechanism \cite{deser}, which has been 
succesfully applied to formulate Yang-Mills, gravity \cite{deser},
supergravity \cite{bdk}. topologically massive Yang-Mills \cite{aragone} 
and Chapline-Manton \cite{aragjor} theories.
 
\section{THE ABELIAN MODEL}

Our starting point will be a first-order formulation for the
Cremmer-Sherk 
theory. This is realized introducing an auxiliary vector field ($v_m$) 
$a la$ Freedman-Townsend. The action is written down as \cite{conv} 
\begin{eqnarray}
I &=& <-\frac{1}{4}\mu\epsilon^{mnpq}B_{mn}[\partial_{p}v_q -
\partial_{q}v_p] 
- \frac{1}{2}\mu^2 v^mv_m -
\frac{1}{2}\mu\epsilon^{mnpq}B_{mn}\partial_{p}A_q\\ 
&+& \frac{1}{4}F_{mn}F^{mn} - \frac{1}{2}F^{mn}[\partial_{m}A_n - 
\partial_{n}A_m]>\nonumber
\end{eqnarray}
where $< >$ denotes integration in four dimensions. All the fields
involved 
have mass dimensions and $\mu$ is a mass parameter. There are two sets 
of abelian gauge invariances:
\begin{equation}
\delta_{\lambda}A_m = \partial_m \lambda, \quad \delta_{\lambda}F_{mn} =
0
\end{equation}
\begin{equation}
\delta_{\zeta}B_{mn} = \partial_m \zeta_n - \partial_n \zeta_m, \quad
\delta_{\zeta}
v_{m} = 0.
\end{equation}

Independent variations in $v_m, B_{mn}, F_{mn}$ and $A_{m}$ lead to the 
following equations of motion
\begin{equation}
v^m = -\frac{1}{6\mu}\epsilon^{mnpq}H_{npq},
\end{equation}
\begin{equation}
\epsilon^{mnpq}\partial_{p}[v_q + A_q] = 0,
\end{equation}
\begin{equation}
F_{mn} = \partial_{m}A_n - \partial_{n}A_m,
\end{equation}
\begin{equation}
\partial_p F^{pm} = \frac{1}{6}\mu\epsilon^{mnpq}H_{npq}
\end{equation}
where $H_{mnp} \equiv \partial_m B_{np} + \partial_n B_{pm} + \partial_p
B_{mn}$ 
is the field strength associated with the antisymmetric field. The
Cremmer-Sherk 
action is obtained after substituing equations (4) and (6) in (1):
\begin{equation}
I_{CrSc} = -\frac{1}{4}F_{mn[A]}F^{mn}_{[A]} -
\frac{1}{12}H_{mnp[B]}H^{mnp}_{[B]} 
- \frac{1}{4}\mu\epsilon^{mnpq}B_{mn}F_{pq[A]}.
\end{equation}

On the other hand, equation (5) can be solved(locally) for the $v$
field,
\begin{equation}
v_{m} = -[A_{m} + \frac{1}{\mu}\partial_{m}\phi],
\end{equation}
where $\phi$ is a scalar field. Substituting this solution 
in the action $I$, the Stuckelberg 
formulation for massive abelian vector bosons is obtained
\begin{equation}
I_{St} = -\frac{1}{4}F_{mn[A]}F^{mn}_{[A]} - \frac{1}{2}\mu^2
[A_{m} + \frac{1}{\mu}\partial_{m}\phi][A^{m} +
\frac{1}{\mu}\partial^{m}\phi].
\end{equation}

As it is well known, both formulations(Stuckelberg and Cremmer-Sherk)
are 
equivalent descriptions of massive abelian gauge invariant vectorial 
theories and propagate three degrees of freedom. This equivalence is 
reflected by the fact that they are connected by duality \cite{siva}.
Indeed, since the 
scalar field appears in equation (10) only through its derivative, we
can apply 
the dualization method due to Nicolai and Townsend \cite{nit}, which
consist
in replacing $\partial_m\phi$ by $\frac{1}{2}l_m$ and adding a 
new term to equation (10): $\epsilon B\partial l$, i.e.
\begin{equation}
I_{Stmod} = -\frac{1}{4}F_{mn[A]}F^{mn}_{[A]} - \frac{1}{2}\mu^2
[A_{m} + \frac{1}{2\mu}l_m][A^{m} + \frac{1}{2\mu}l_m] + 
\frac{1}{4}\epsilon^{mnpq}B_{mn}\partial_{p}l_q.
\end{equation}

At this stage, $B_{mn}$ is a Lagrange multiplier forcing the constraint 
$\partial_m l_n - \partial_n l_m = 0$ whose local solution is $l_m = 
2\partial_m\phi$. Now, if we eliminate $l_m$ via its equation of 
motion
\begin{equation}
l^m = \frac{1}{3}\epsilon^{mnpq}H_{npq} -2\mu A^m
\end{equation}
and go back to equation (11), the Cremmer-Sherk action is recovered.

Finally, let us recall that the second-order field equations can be
written as
\begin{equation}
\partial_p F^{pm} = J^m, \quad \partial_p H^{pmn} = J^{mn},
\end{equation}
where 
\begin{equation}
J^{m} = \frac{1}{6}\mu\epsilon^{mnpq}H_{npq} \quad \mbox{and} \quad
J^{mn} = 
\frac{1}{2}\mu\epsilon^{mnpq}F_{pq}
\end{equation}
are "topological"currents in the sense that they are conserved without 
using the equations of motion.

\section{THE SELF-INTERACTION PROCESS}

Now, we extend the first-order action, equation (1), by introducing a 
triplet of free abelian antisymmetric fields $B_{mn}^a$ coupled with a
triplet 
of free abelian vector fields $A_{m}^a$, ($a = 1,2,3$)

\begin{eqnarray}
I_o &=& <-\frac{1}{4}\mu\epsilon^{mnpq}B_{mn}^a[\partial_{p}v_q^a 
- \partial_{q}v_p^a] 
- \frac{1}{2}\mu^2 v^{am}v_m^a - 
\frac{1}{2}\mu\epsilon^{mnpq}B_{mn}^a\partial_{p}A_q^a\\ 
&+& \frac{1}{4}F_{mn}^aF^{amn} - \frac{1}{2}F^{amn}[\partial_{m}A_n^a - 
\partial_{n}A_m^a]>\nonumber
\end{eqnarray}

Besides the local gauge transformations
\begin{equation}
\delta_{\lambda}A_m^a = \partial_m \lambda^a, \quad
\delta_{\lambda}F_{mn}^a = 0
\end{equation}
\begin{equation}
\delta_{\zeta}B_{mn}^a = \partial_m \zeta_n^a - \partial_n \zeta_m^a,
 \quad \delta_{\zeta}v_{m}^a = 0,
\end{equation}
our action has two global 
invariances: one is a global $SU(2)$ rotation and the other is a 
a global symmetry associated with the Freedman-Townsend theory:
\begin{equation}
(I) \quad \delta_{\omega}X^a = g_1\epsilon^{abc}X^b\omega^c
\end{equation}
where $X^a = (A_m^a,F_{mn}^a,v_m^a,B_{mn}^a)$ and
\begin{eqnarray}
(II) \quad \delta_{\rho}B_{mn}^a &=& g_2\epsilon^{abc}[v_m^b +
A_m^b]\rho_n^c 
- m\leftrightarrow n,\\
 \delta_{\rho}v_m^a &=& \delta_{\rho}A_m^a =\delta_{\rho}F_{mn}^a =
0,\nonumber
\end{eqnarray}
$\omega$ and $\rho$ being global parameters. 
In principle the coupling constants $g_1$ and $g_2$ are different. We
note 
that under type II transformations the action changes by a total
derivative. 
The Noether currents associated to these invariances are given by
\begin{equation}
g_1^{-1}j^{am} = \epsilon^{abc}F^{bmn}A_n^c +
\frac{1}{2}\mu\epsilon^{mnpq}
\epsilon^{abc}B_{pq}^b[A_n^c + v_n^c]
\end{equation}
and
\begin{equation}
g_2^{-1}K^{amn} = \frac{1}{2}\mu\epsilon^{mnpq}\epsilon^{abc}[A_p^b +
v_p^b]
[A_q^c + v_q^c].
\end{equation}

These are conserved on-shell. In order to couple these 
currents to the action $I_o$ we must add the corresponding
self-interaction 
terms: $I_1$ and $I_2$ defined by:
\begin{equation}
j^{am} \equiv \frac{\delta I_1}{\delta_{A_m^a}}; \quad 
K^{amn} \equiv -2\frac{\delta I_2}{\delta_{B_{mn}^a}}.
\end{equation}

These functional differential equations can easily be integrated.
In fact, we find that
\begin{eqnarray}
I_1 &=& -g_1 <\frac{1}{2}\epsilon^{abc}F^{amn}A_m^bA_n^c +
\frac{1}{4}\mu
\epsilon^{mnpq}\epsilon^{abc}B_{mn}^aA_p^bA_q^c \\
&+&
\frac{1}{2}\mu\epsilon^{mnpq}\epsilon^{abc}B_{mn}^aA_p^bv_q^c>\nonumber
\end{eqnarray}
and
\begin{eqnarray}
I_2 &=& -g_2 <\frac{1}{4}\epsilon^{mnpq}\epsilon^{abc}B^{amn}v_p^bv_q^c 
+ \frac{1}{4}\mu\epsilon^{mnpq}\epsilon^{abc}B_{mn}^aA_p^bA_q^c \\
 &+&
\frac{1}{2}\mu\epsilon^{mnpq}\epsilon^{abc}B_{mn}^aA_p^bv_q^c>\nonumber
\end{eqnarray}

However, these two terms have overlapping parts. This situation is akin 
to what happens in the derivation of supergravity from 
self-interaction \cite{bdk}. In order to overcome this obstacle we 
must require equality of the coupling constants: $g \equiv  g_1 = g_2$
and write down the self-interaction action as

\begin{eqnarray}
I_{SI} &\equiv& -g< \frac{1}{2}\epsilon^{abc}F^{amn}A_m^bA_n^c 
+ \frac{1}{4}\epsilon^{mnpq}\epsilon^{abc}B_{mn}^{a}v_p^bv_q^c \\
&+& \frac{1}{4}\mu\epsilon^{mnpq}\epsilon^{abc}B_{mn}^aA_p^bA_q^c
+
\frac{1}{2}\mu\epsilon^{mnpq}\epsilon^{abc}B_{mn}^aA_p^bv_q^c>\nonumber
\end{eqnarray}

Actually, we have that 
\begin{equation}
j^{am} \equiv \frac{\delta I_{SI}}{\delta_{A_m^a}}\quad \mbox{and} \quad 
K^{amn} \equiv -2\frac{\delta I_{SI}}{\delta_{B_{mn}^a}}.
\end{equation}

The self-interaction mechanism stops here since no other 
derivative terms appear in $I_{SI}$. Finally, the full non-abelian
theory is

\begin{eqnarray}
I &=& I_o + I_{SI}\\
&=& <- \frac{1}{4}\mu\epsilon^{mnpq}B_{mn}^a[F_{pq}^a + f_{pq}^a +
2\epsilon^{abc}
A_p^bv_p^c] - \frac{1}{2}\mu^2 v_m^av^am - \frac{1}{4}F_{mn}^aF^{amn}
>\nonumber,
\end{eqnarray}
where 
\begin{equation}
F_{mn}^a \equiv \partial_mA_n^a - \partial_nA_m^a +
g\epsilon^{abc}A_m^bA_n^c
\end{equation}
and
\begin{equation}
f_{mn}^a \equiv \partial_mv_n^a - \partial_nv_m^a +
g\epsilon^{abc}v_m^bv_n^c
\end{equation}
which is just that proposed by Freedman and Townsend (equation (2.15) in 
their paper). As usual, the self-interaction process combines the
abelian gauge 
transformations with the global ones giving rise to non-abelian local
gauge 
transformations. In our case, we have 
\begin{eqnarray}
\delta_{\alpha}A_m^a &=& \partial_m\alpha^a +
g\epsilon^{abc}A_m^b\alpha^c\\
\delta_{\alpha}B_{mn}^a &=& g\epsilon^{abc}B_{mn}^b\alpha^c\\
\delta_{\alpha}v_m^a &=& g\epsilon^{abc}v_m^b\alpha^c\nonumber
\end{eqnarray}
and
\begin{eqnarray}
\delta_{\xi}B_{mm}^a &=& \partial_m\xi^a + g\epsilon^{abc}[A_m^b +
v_m^b] 
\xi^c - m\leftrightarrow n\\
\delta_{\xi}A_m^a &=& 0 = \delta_{\xi}v_m^a\nonumber.
\end{eqnarray}
The action of Freedman-Townsend, equation (27), is equivalent to 
massive Yang-Mills (locally) 
as can be shown after elimination of $B_{mn}^a$ through its equation of
motion, 
which said us that $A_m + v_m$ is a pure gauge.

\section{CONCLUSION}
In this letter, by starting with a nice abelian first-order formulation,
and through the application of the self-interaction mechanism 
we have obtained the Freedman-Townsend theory and 
its corresponding gauge tranformation rules through self-interaction.
The first order abelian formulation allowed us to find Cremmer-Sherk 
and Stuckelberg formulations for massive spin-1 theories, these  
later formultations are connected by duality. The BRST quantization 
of the massive Freedman-Townsend has been performed 
by Thierry-Meig \cite{tm}. Since massive Freedman-Townsend theory is
equivalent 
(in topologically trivial manifols) to massive Yang Mills it should be 
interesting to attempt to connect Friedman-Townsend with others
approaches 
dealing with massive 
gauge bosons without the presence of Higgs field \cite{vani}.  

\section{ACKNOWLEDGEMENT}
I thank P.J. Arias for useful
discussions and U. Percoco, M. Caicedo, 
N. Pantoja and L. Labrador also for carefully reading the manuscript.


\begin{thebibliography} {60}
\bibitem{conv} We use the metric $\eta_{mn} = diag(-1,+1,+1,+1)$.
\end{thebibliography}
\end{document}